\overfullrule=0pt

\input harvmac

\def\a{\alpha}
\def\ad{{\dot\alpha}}
\def\ah{{\widehat\alpha}}
\def\b{\beta}
\def\bd{{\dot\beta}}
\def\bh{{\widehat\beta}}
\def\g{\gamma}

\def\l{\lambda}

\def\lb{\overline\lambda}
\def\d{\delta}
\def\t{\theta}
\def\tt{\overline\theta}
\def\pp{\overline p}
\def\dd{\overline d}

\def\s{\sigma}
\def\o{\omega}

\def\ob{\overline\omega}
\def\O{\Omega}

\def\L{\Lambda}

\def\Pb{\overline\Pi}

\def\p{\partial}
\def\pb{\overline\partial}

\def\half{{1\over 2}}
\def\mh{{\widehat m}}
\def\nh{{\widehat n}}
\def\ph{{\widehat p}}
\def\qh{{\widehat q}}
\def\e{\epsilon}
\def\YB{\overline Y}
\def\yb{\overline y}
\def\vv{\overline v}

\Title{ \vbox{}} {\vbox{\centerline{ D=4 Pure Spinor Superstring
and N=2 Strings}}}

\smallskip
\centerline{Osvaldo Chand\'{\i}a\foot{e-mail: ochandia@unab.cl}}
\smallskip
\centerline{\it Departamento de Ciencias F\'{\i}sicas, Universidad
Andr\'es Bello } \centerline{\it Rep\'ublica 252, Santiago, Chile}

\bigskip

\noindent We study the compactification of the pure spinor
superstring down to four dimensions. We find that the compactified
string is described by a conformal invariant system for both the
four dimensional and for the compact six dimensional variables.
The four dimensional sector is found to be invariant under a
non-critical N=2 superconformal transformations.

\Date{September 2005}

\newsec{Introduction}

The pure spinor formalism of the superstring has solved the
longstanding problem of the covariant quantization of the
superstring \ref\BerkovitsFE{N.~Berkovits, ``Super-Poincare
Covariant Quantization of the Superstring,'' JHEP 0004 (2000) 018
[arXiv:hep-th/0001035].}. This formalism has passed many tests since
it was formulated. It describes correctly the superstring spectrum
in the light cone gauge \ref\BerkovitsNN{N.~Berkovits, ``Cohomology
in the Pure Spinor Formalism for the Superstring,'' JHEP 0009 (2000)
046 [arXiv:hep-th/0006003]\semi N.~Berkovits and O.~Chand\'{\i}a,
``Lorentz Invariance of the Pure Spinor BRST Cohomology for the
Superstring,'' Phys.\ Lett.\ B514 (2001) 394
[arXiv:hep-th/0105149].}, in semi light cone gauge
\ref\BerkovitsTW{N.~Berkovits and D.~Z.~Marchioro, ``Relating the
Green-Schwarz and Pure Spinor Formalisms for the Superstring,'' JHEP
0501 (2005) 018 [arXiv:hep-th/0412198].} (see also
\ref\GaonaYW{A.~Gaona and J.~A.~Garcia, ``BFT Embedding of the
Green-Schwarz Superstring and the Pure Spinor Formalism,''
[arXiv:hep-th/0507076].}) and in a manifestly ten dimensional
covariant manner \ref\BerkovitsQX{N.~Berkovits and O.~Chand\'{\i}a,
``Massive Superstring Vertex Operator in D = 10 Superspace,'' JHEP
0208 (2002) 040 [arXiv:hep-th/0204121].}. It has also been possible
to compute amplitudes by using this formalism. Tree-level amplitudes
were shown to coincide with the RNS result in
\ref\BerkovitsPH{N.~Berkovits and B.~C.~Vallilo, ``Consistency of
Super-Poincare Covariant Superstring Tree Amplitudes,'' JHEP 0007
(2000) 015 [arXiv:hep-th/0004171].} and loop amplitudes were defined
and used to prove certain non renormalization theorems for low
energy effective action terms in \ref\BerkovitsPX{N.~Berkovits,
``Multiloop Amplitudes and Vanishing Theorems Using the Pure Spinor
Formalism for the Superstring,'' JHEP 0409 (2004) 047
[arXiv:hep-th/0406055].}.

The formalism has been used to construct quantizable sigma model
actions in curved backgrounds. The action for a generic
supergravity/super Yang-Mills background was studied in
\ref\BerkovitsUE{N.~Berkovits and P.~S.~Howe, ``Ten-dimensional
Supergravity Constraints from the Pure Spinor Formalism for the
Superstring,'' Nucl.\ Phys.\ B635 (2002) 75
[arXiv:hep-th/0112160].} at the classical level and its conformal
invariance at the quantum level was verified in
\ref\ChandiaHN{O.~Chand\'{\i}a and B.~C.~Vallilo, ``Conformal
Invariance of the Pure Spinor Superstring in a Curved
Background,'' JHEP 0404 (2004) 041 [arXiv:hep-th/0401226].}.
Backgrounds supporting Ramond Ramond fluxes can also be
constructed in this framework. Namely, the AdS$_5\times$S$^5$ case
was classically studied in \ref\BerkovitsYR{N.~Berkovits and
O.~Chand\'{\i}a, ``Superstring Vertex Operators in an
AdS$_5\times$S$^5$ Background,'' Nucl.\ Phys.\ B596 (2001) 185
[arXiv:hep-th/0009168].} and quantum mechanically in
\ref\ValliloMH{B.~C.~Vallilo, ``One Loop Conformal Invariance of
the Superstring in an  AdS$_5\times$S$^5$ Background,'' JHEP 0212
(2002) 042 [arXiv:hep-th/0210064]\semi N.~Berkovits, ``Quantum
Consistency of the Superstring in AdS$_5\times$S$^5$ Background,''
JHEP 0503 (2005) 041 [arXiv:hep-th/0411170].}.

Attempts to extract a manifest space time supersymmetry from the
RNS formalism is only possible if the string is compactified to
four \ref\BerkovitsBF{N.~Berkovits, ``A New Description of the
Superstring,'' arXiv:hep-th/9604123\semi N.~Berkovits,
M.~Bershadsky, T.~Hauer, S.~Zhukov and B.~Zwiebach, ``Superstring
Theory on AdS$_2\times$S$^2$ as a Coset Supermanifold,'' Nucl.\
Phys.\ B567 (2000) 61 [arXiv:hep-th/9907200].} or to six
dimensions \ref\BerkovitsIM{N.~Berkovits, C.~Vafa and E.~Witten,
``Conformal Field Theory of AdS Background with Ramond-Ramond
Flux,'' JHEP 9903 (1999) 018 [arXiv:hep-th/9902098]\semi
N.~Berkovits, ``Quantization of the Type II Superstring in a
Curved Six-dimensional Background,'' Nucl.\ Phys.\ B565 (2000) 333
[arXiv:hep-th/9908041].} . In ten dimensions is possible to
preserve a U(5) subgroup of SO(10) \ref\BerkovitsIN{N.~Berkovits,
``Quantization of the Superstring with Manifest U(5)
Super-Poincare Invariance,'' Phys.\ Lett.\ B457 (1999) 94
[arXiv:hep-th/9902099].}. Since the pure spinor formalism is
manifestly supersymmetric in ten dimensions, it is tempting to
find a suitable compactification and relate the resulting theory
to a hybrid string. We pursue this objective for the four
dimensional version of the hybrid superstring. We will find that
the four dimensional pure spinor string has the same matter
world-sheet variables as the hybrid superstring. They differ in
the ghosts variables. In the pure spinor case, the ghosts are
given by an unconstrained c-number Weyl spinor, while for the
hybrid string the ghost is a chiral boson. As for the hybrid
string, our compactified pure spinor string will be invariant
under N=2 superconformal transformations but the corresponding
algebra has a different central charge. During the conclusion of
this work, we have noticed the appearance of the papers
\ref\BerkovitsBT{N.~Berkovits, ``Pure Spinor Formalism as an N=2
Topological String,'' [arXiv:hep-th/0509120].},
\ref\GrassiSB{P.~A.~Grassi, N.~Wyllard, ``Lower-dimensional
Pure-spinor Superstrings,'' [arXiv:hep-th/0509140].} and
\ref\CategoriesFH{N.~Wyllard, ``Pure-spinor Superstrings in
d=2,4,6,'' [arXiv:hep-th/0509165].} with some overlapping with our
results.

In the next section we will compactify the ten dimensional pure
spinor string to preserve four dimensional supersymmetry. In
section 3 we will find an N=2 superconformal structure, although
the central charge turns out to be $c=-6$ for the four dimensional
sector.

\newsec{The Four Dimensional Pure Spinor Superstring}

In this section we will compactify the ten dimensional pure spinor
superstring. We will first review the ten dimensional pure spinor
string and then we will compactify down to four dimensions.

\subsec{Review of the ten dimensional pure spinor string}

The pure spinor string in ten dimensions is described by the
free-field conformal invariant action

\eqn\spure{S = \int d^2z \half \p X^\mh \pb X_\mh + p_\ah \pb
\t^\ah + S_{pure},} where $(X^\mh, \t^\ah)$ with $\mh=0,\dots, 9$
and $\ah=1,\dots, 16$ define the ten-dimensional superspace,
$p_\ah$ is the conjugate momentum of $\t^\ah$ and $S_{pure}$ is
the action for the pure spinor ghosts variables. They are a
c-number ten dimensional spinor $\l^\ah$ constrained by the
condition $(\l\g^\mh\l)=0$ and its conjugate $\o_\ah$ which is
defined up to the gauge transformation $\d\o_\ah = \L_\mh (\g^\mh
\l)_\ah$, here the $\g^\mh$ are the symmetric $16\times 16$
ten-dimensional Pauli matrices. One can check that the pure spinor
constraint implies that only 11 out of 16 components of $\l^\ah$
are independent. Similarly the gauge transformation for $\o_\ah$
implies that 11 out of its sixteen components are independent.
Therefore, the action of \spure\ is conformal invariant. That is,
the central charge vanishes since we have 32 bosons and 32
fermions in the world-sheet action of \spure.

Physical states are defined as ghost number one vertex operators
in the cohomology of the BRST operator $Q=\oint \l^\ah d_\ah$,
where

$$d_\ah = p_\ah - \half (\g^\mh \t)_\ah \p X_\mh -{1\over 8} (\g^\mh
\t)_\ah (\t \g_\mh \p \t),$$ is the world-sheet generator of
superspace translations. Since the $d_\ah$'s satisfy the OPE

$$
d_\ah(y) d_\bh(z) \to -{1\over(y-z)} \g^\mh_{\ah\bh} \Pi_\mh(z),$$
where $\Pi^\mh=\p X^\mh + \half (\t \g^\mh \p \t )$, it is trivial
to check the nilpotence of the BRST  charge. A physical state is
given by a vertex operator $V$ which is annihilated by $Q$ up to an
operator of the form $Q\O$. At ghost number one, massless states of
the open strings are shown to describe super Maxwell multiplet in
ten dimensions after using BRST conditions. Namely, a ghost number
one vertex operator is of the form $V = \l^\ah A_\ah(X,\t)$, where
$A_\ah(X,\t)$ is superfield which depends on the zero modes of
$(X,\t)$ only. It guarantees that $V$ is describing massless fields.
It is easy to obtain the equation of motion derived by the condition
$QV=0$. It is $\l^\ah \l^\bh D_\ah A_\bh=0$, where $D_\ah$ is the
space time supersymmetric covariant derivative. This equation is
solved by $D_{(\ah} A_{\bh)}=\g^\mh_{\ah\bh} A_\mh$. The gauge
invariance is obtained from the BRST invariance $\d V= Q\O$ which
implies $\d A_\ah = D_\ah \O$. After using this gauge invariance one
can show that the $\t$ independent part of the superfield $A_\mh$ is
the photon and the term linear in $\t$ involves the photino.

So far, we have no discussed the pure spinor constraint. We have
only used its main property to check the nilpotence of the BRST
charge. In order to verify Lorentz invariance, to construct massive
states and to properly define loop amplitudes, we need to solve the
pure spinor constraint. The pure spinor ghosts can only appear in
combinations that preserve the gauge invariance $\d \o_\ah = \L_\mh
(\g^\mh \l)_\ah$. At ghost number zero, the only possible choices
are $\o_\ah \p \l^\ah, \l^\ah \o_\ah$ and $(\o \g^{\mh\nh} \l)$.
They are respectively the stress tensor $T_\l$, the ghost number
current $J$ and the Lorentz current $N^{\mh\nh}$ for the pure spinor
ghosts. In order to compute the algebras for these generators one
has to temporally break Lorentz covariance. For example one can
choose to preserve a U(5) subgroup of SO(10) as in \BerkovitsFE.
Although the computation is not covariant, the result is. The OPE
algebra one obtains is

\eqn\curr{N^{\mh\nh}(y) N^{\ph\qh}(z) \to {3\over(y-z)^2}
\eta^{\mh[\ph} \eta^{\qh]\nh} + {1\over(y-z)} ( \eta^{\mh[\ph}
N^{\qh]\nh} - \eta^{\nh[\ph} N^{\qh]\mh} ),}
$$
J(y) J(z) \to -{4\over(y-z)^2},\quad T_\l(y) T_\l(z) \to
{22\over{2(y-z)^4}} + {{2T_\l(z)}\over(y-z)^2} + {{\p
T_\l(z)}\over(y-z)},$$
$$
T_\l(y) J(z) \to {8\over(y-z)^3} + {J(z)\over(y-z)^2} + {{\p
J(z)}\over(y-z)}.$$ This algebra determine the current $N^{\mh\nh}$
to be the Lorentz generator for the pure spinor variables and give
rise to the right double pole coefficient in the $N^{\mh\nh}(y)
N^{\ph\qh}(z)$ OPE. The central charge is $22$ which cancel the
$-22$ contribution from the matter variables and leads to an anomaly
of $+8$ in the ghost number current.

\subsec{Compactification to four dimensions}

Now we go to four dimensions. First of all, we need to write the
SO(10) representations such that we can extract SO(4)
representations. By using the SU(4) notation of
\ref\GrassiXC{P.~A.~Grassi and P.~van Nieuwenhuizen, ``Harmonic
Superspaces from Superstrings,'' Phys.\ Lett.\ B593 (2004) 271
[arXiv:hep-th/0402189].}, the pure spinor $\l^\ah$ can be
decomposed as

\eqn\lsufour{\l^\ah = (\l^\a_A, \lb^{\ad A}),} where $\a, \ad = 1,
2, A=1,\dots, 4.$ The pure spinor condition $(\l\g^{\mh}\l)=0$
decomposes in the following conditions

\eqn\pcsufour{\l^\a_A \lb^{\ad A}=0,\quad \e_{\a\b} \l^\a_A
\l^\b_B + \half \e_{\ad\bd} \e_{ABCD} \lb^{\ad C} \lb^{\bd D} =0,}
the first comes from $\mh=0,\dots, 3$ and the second from the
remaining directions.

To extract a four dimensional spinor, we make one more
decomposition. We use SU(3)$\times$U(1) instead of SU(4) to write

\eqn\lsuthree{\l^\a_A = ( \l^\a, \l^\a_I ),\quad \lb^{\ad A} = (
\lb^{\ad}, \lb^{\ad I} ),} where $I=1,\dots, 3$. The non trivial
pure spinor conditions in \pcsufour\ become

\eqn\pcsuthree{\l^\a \lb^\ad + \l^\a_I \lb^{\ad I} = 0,}
$$
\e_{\a\b} \l^\a \l^\b_I + \half \e_{\ad\bd} \e_{IJK} \lb^{\ad J}
\lb^{\bd K} = 0,$$
$$
\e_{\a\b} \l^\a_I \l^\b_J + \e_{\ad\bd} \e_{IJK} \lb^\ad \lb^{\bd
K} = 0.$$ Since pure spinors in four dimensions have only two
components \ref\BerkovitsHY{N.~Berkovits and N.~Nekrasov, ``The
Character of Pure Spinors,'' [arXiv:hep-th/0503075].}, we expect
that a non trivial solution to these conditions is

\eqn\lsol{\l^\a_A = ( \l^\a, \l^\a_I  ),\quad \lb^{\ad A} = ( 0 ,
\lb^{\ad I} ).} The field $\l^\a$ is unconstrained and the
remaining components are constrained according to \pcsuthree.
After putting $\lb^\ad=0$ in the third condition in \pcsuthree\
one obtains

\eqn\sl{\e_{\a\b} \l^\a_I \l^\b_J = 0,} which implies $\l^2_I$ is
proportional to $\l^1_I$ (it will be called as $\l_I$). Then we
have 11 components for a pure spinor, namely $( \l^\a, \l_I,
\lb^{\ad I})$.

Now we examine if we can gauge fix some components of the
conjugate pure spinor variable $\o_\ah$. Recall that it is defined
up to the gauge transformation $\d \o_\ah = \L_\mh (\g^\mh
\l)_\ah$. We note first that only five of the ten gauge parameters
$\L_\mh$ are independent. We write $\L_\mh = ( \L_m, \L_I, \L_{IJ}
)$, where $\L_{IJ}$ is antisymmetric. Only two of the four $\L_m$
are independent and only three of the six $( \L_I, \L_{IJ} )$,
here we can choose $\L_{IJ}=0$. Using this gauge invariance we can
gauge fix to zero the components $\ob_\ad$ and $\o_2^I$. In fact,

$$
\d\ob_\ad = \L_m \s^m_{\b\ad} \l^\b,\quad \d\o_\a^I = \L_m
\s^m_{\a\bd} \lb^{\bd I} - \e_{\a\b} \L^I \l^\b,$$ by using the two
independent $\L_m$ to gauge fix $\ob_\ad$ and the three $\L^I$ to
gauge fix the three $\o^I_2$.

The BRST current can be written as

$$
j_{BRST} =\l^\ah d_\ah = \l^\a d_\a + \l^\a_I d^I_\a + \lb^{\ad I}
\dd_{\ad I} ,$$ where the ten-dimensional world-sheet generator of
superspace translations $d_\ah$ is decomposed as $(d_\a, \dd_\ad,
d_\a^I, \dd_{\ad I})$ and $\l^\a_I$ constrained as \sl. Physical
states are in the cohomology of $Q = \oint j_{BRST}$, for those
states which are independent of the compactification variables the
BRST current is simply equal to $\l^\a d_\a$. As was noted in
\BerkovitsBT, $d_\a$ can be mapped to $p_\a$. Therefore, the
physical states depend on $(X^m, \tt, {\overline p})$ only and
describe the chiral sector of the open superstring field theory of
\BerkovitsBT.

\newsec{N=2 Superconformal Invariance}

In this section we will study the N=2 superconformal invariance of
the four dimensional part of the pure spinor superstring action of
\spure\ when it is compactified to four dimensions. In the
previous section we solved the pure spinor constraint in a form
which is suitable for the compactification we want to perform. It
remains what to do with the matter part of the action. It is
natural to split $X^\mh = ( X^m , Y^I , \YB^I )$ with $m =
0,\dots, 3$ and $I=1,\dots 3$. For the superspace variables we use
the SU(3)$\times$ U(1) notation as in the previous section. The
action becomes

\eqn\sfour{S = \int d^2z \half \p X^m \pb X_m + p_\a \pb \t^\a +
\pp_\ad \pb \tt^\ad + \o_\a \pb \l^\a + S_C,} where $S_C$ stands
for the compact part of the action and it is given by

\eqn\sc{S_C = \int d^2z ~ \p \YB^I \pb Y_I + p_\a^I \pb \t^\a_I +
\pp_\ad^I \pb \tt^\ad_I + \ob_{\ad I} \pb \lb^{\ad I} + \o^I_\a \pb
\l_I^\a,} where $\l^\a_I$ is constrained according to \sl. We note
that our action has vanishing central charge. Now it will be proven
that both the four dimensional and the compact six dimensional part
of the action have N=2 superconformal invariance.

\subsec{N=2 D=4}

For the four dimensional part of the action we have eight bosons
and eight fermions, then the stress tensor becomes

$$
T = -\half \p X^m \p X_m - p_\a \p \t^\a - \pp_\ad \p \tt^\ad -
\o_\a \p \l^\a,$$ and satisfies the OPE algebra

$$
T(y) T(z) \to {{2T(z)}\over(y-z)^2} + {{\p T(z)}\over(y-z)}.$$

In order to relate the pure spinor string to the hybrid formalism,
one needs to show a hidden N=2 superconformal invariance. One is
tempted to write the twisted N=2 generators to be

\eqn\ntwo{T = -\half \p X^m \p X_m - p_\a \p \t^\a - \pp_\ad \p
\tt^\ad - \o_\a \p \l^\a,}
$$
G^+ = \l^\a d_\a,$$
$$
G^- = \o_\a \p \t^\a + \Pi^m (y \s_m \dd),$$
$$
J = \l^\a \o_\a,$$ where $\Pi^m$ is the four dimensional projection
of $\Pi^\mh$ and $d_\a, \dd_\ad$ come from the ten dimensional field
$d_\ah$. The field $y_\a$ comes from the ten dimensional projector
field $y_\ah$ defined as $y_\ah = v_\ah/(v_\bh \l^\bh)$ with the
property $y_\ah \l^\ah=1$ \ref\MatoneFT{M.~Matone, L.~Mazzucato,
I.~Oda, D.~Sorokin and M.~Tonin, ``The Superembedding Origin of the
Berkovits Pure Spinor Covariant Quantization of Superstrings,''
Nucl.\ Phys.\ B639 (2002) 182 [arXiv:hep-th/0206104].}. Since
$y_\ah$ turns out to be a pure spinor, one can use the decomposition
of the previous section to consider only $y_\a$ defined as $v_\a /
(v_\b \l^\b)$ with $v_\a$ an arbitrary but constant c-number spinor
in four dimensions and constrained by $\l^\a y_\a = 1$. In order to
verify that the generators \ntwo\ satisfy the right N=2 algebra one
has to check that the OPE's $G^+(z) G^+(w)$ and $G^-(z) G^-(w)$ are
not singular. It is easy to verify this for $G^+$. For $G^-$ one has
to be careful. There is a potential singular term coming from the
contraction between $\Pi^m$ with itself which has the form

$$
{1\over(z-w)} \e^{\a\b} y_\a \p y_\b \dd_\ad \dd^\ad.$$ Recall the
definition of the projector field $y_\a=v_\a / (v_\b \l^\b)$ with
$v_\a$ being a constant c-number spinor \MatoneFT. From this
definition one obtains

$$
\p y_\a = -y_\a y_\b \p \l^\b,$$ plugging this into the above
equation one see that the potential singular term is zero since
$\e^{\a\b} y_\a y_\b =0$.

This construction failed to be a critical system since the $J(y)
J(z)$ OPE leads to a central charge of ${\hat c}=-2$ instead of
the critical value of $+2$. A solution for this problem is the non
minimal construction of \BerkovitsBT.

\subsec{N=2 D=6}

For the compact six dimensional part of the action, we must to
take into account the constraint \sl\ for the pure spinor
component $\l^\a_I$. The N=2 generators are

\eqn\ntwoc{T_C = -\p Y^I \p \YB_I -p_\a^I \p \t^\a_I - \pp_{\ad I}
\p \tt^{\ad I} + \ob_{\ad I} \p \lb^{\ad I} + \o^I_\a \p \l^\a_I,}
$$
G^+_C = \lb^{\ad I} d_{\a I} + \l^\a_I d_\a^I,$$
$$
G^-_C = \ob_{\ad I} \p \tt^{\ad I} + \o_\a^I \p \t^\a_I + \e_{IJK}
 \e^{\a\b} \Pi^I y_\a^J d_\b^K + \e^{IJK} \e^{\ad\bd} \Pb_I  \yb_{\ad
I} \dd_{\bd K}  ,$$
$$
J_C = \lb^{\ad I} \ob_{\ad I} + \l^\a_I \o_\a^I,$$ where $\Pi^I$
and $\Pb_I$ are the projections of $\Pi^\mh$ to six dimensions and
both have conformal weight of $(1,0)$ . As in the four dimensional
case, we need to introduce a projector field $(y_\a^I, \yb_{\ad
I})$ defined by

$$
y_\a^I = {{ v_\a^I}\over{(\l^\b_J v_\b^J)}},\quad \yb_{\ad I} =
{{\vv_{\ad I}\over{(\lb^{\bd J} \vv_{\bd J})}}},$$ with the
c-number spinors $v_\a^I$ and $\vv_{\ad I}$ being arbitrary
constants. The projectors are constrained to satisfy

$$\l^\a_I y_\a^I = \lb^{\ad I} \yb_{\ad I} = 1.$$
As in four dimensions, the hard to check part of the algebra is
the OPE $G^-_C(z) G^+_C(w)$ which again vanishes in virtue of the
properties of the projector fields. The N=2 superalgebra has the
central charge of $\hat{c}=-9$. One can obtain this result by
temporarily break Lorentz invariance such that the only physical
component for the pure spinor ghosts are $\lb^{\ad I}$ and one of
the spinor components of $\l^\a_I$ (say $\l^1_I \equiv \l_I$) and
their conjugates variables. Therefore, $J_C = \lb^{\ad I} \ob_{\ad
I }+ \l_I \o^I$, then $J_C(z) J_C(w) \to -9(z-w)^{-2}$.

\vskip 15pt {\bf Acknowledgements:} I would like to thank Nathan
Berkovits,  William Linch and Brenno Vallilo for useful comments
and suggestions. I would also like to thank the participants and
organizers of the Simons Workshop in Mathematics and Physics where
part of this work was done. Partial financial support from
Fundaci\'on Andes is also acknowledged.

\listrefs

\end